\documentclass[aps,nopacs,showkeys,nofootinbib,preprint]{revtex4}
\usepackage{epsfig,amssymb,bm,graphicx,color,amsmath,rotating}
\usepackage{xcolor}
\usepackage[percent]{overpic}
\usepackage{mwe,xcolor}
\usepackage{transparent}
\usepackage{amsmath}
\usepackage{framed}
\usepackage{slashed}
\usepackage{diagbox}
\usepackage{sidecap}
\usepackage{hyperref}
\newcommand\arx[3][r]{
  \ifx r#1 \href{https://arxiv.org/abs/#2}{[arXiv:#2].} \else
  \ifx o#1 \href{https://arxiv.org/abs/#3/#2}{[arXiv:#3/#2].} \else
  \ifx b#1 \href{https://arxiv.org/abs/#2}{[arXiv:#2 [#3]].} \else
  {Illegal~option}
  \fi\fi\fi
}
\begin{document} {\normalsize} 

\title{
Core-corona decomposition of compact (neutron) stars compared to NICER data
including  XTE J1814-338
} 

\author{R.~Z\"ollner${}^1$, B.~K\"ampfer${}^{2, \, 3}$}
\affiliation{${}^1$Institut f\"ur Technische Logistik und Arbeitssysteme, TU~Dresden, 01062 Dresden, Germany}
\affiliation{${}^2$Helmholtz-Zentrum  Dresden-Rossendorf, 01314 Dresden, Germany}
\affiliation{${}^3$Institut f\"ur Theoretische Physik, TU~Dresden, 01062 Dresden, Germany}
 
\begin{abstract}
A core-corona decomposition of compact (neutron) star models is compared to recent NICER data
of masses and radii. It is in particular interesting to capture the outlier XTE~J1814-338.
Instead of integrating the TOV equations from the center to surface, 
we follow here another pathway by accommodating all uncertainties of the equation(s) of state (EoS)
at supra-nuclear density or/and an unknown dark matter admixture in a parameterization of the core
by its radius $r_x$, the included mass $m_x$ and the pressure $p_x$ at $r_x$.
The corona, which may be dubbed also envelope or halo or outer crust, is assumed to be of standard-model  matter
where the EoS is supposed to be faithfully known. 
\end{abstract}


\date{\today}

\maketitle

\section{Introduction} \label{introduction}

The advent of detecting gravitational waves from merging neutron stars,
the related multimessenger astrophysics 
and the improving mass-radius determinations
of neutron stars, in particular by NICER data, stimulated a wealth of activities
\cite{Riley:2021pdl,Miller:2021qha,Miller:2019cac,Riley:2019yda,Raaijmakers:2021uju,Kini:2024ggu,Doroshenko:2022nwp,Choudhury:2024xbk,Pang:2021jta,Vinciguerra:2023qxq,Fonseca:2021wxt,Romani:2022jhd,Salmi:2024bss,Salmi:2024aum}.
Besides masses and radii, moments of inertia
and tidal deformabilities become experimentally accessible and can be confronted with theoretical models
\cite{Rutherford:2024srk}.
The baseline of the latter ones is provided by non-rotating, spherically symmetric
cold dense matter configurations. The sequence of white dwarfs (first island of stability)
and neutron stars (second island of stability) and possibly 
a third island of stability 
shows up thereby when going to more compact objects, with details depending sensitively
on the actual equation of state (EoS). 
Since the radii of configurations
of the second (neutron stars) and third (hypothetical quark/hybrid stars) 
islands can be very similar, 
the notion of twin stars 
has been coined for equal-mass configurations. 
These issues and their background are surveyed in \cite{Schaffner-Bielich:2020psc}.

We emphasize the relation of ultra-relativistic heavy-ion collision physics, probing
the EoS at large temperatures and small net-densities, and compact star physics, probing
small temperatures and large net baryon densities when focusing on static compact-star properties.
(Of course, in binary or ternary compact-star merging-events, 
also finite temperatures and a large range of densities are probed 
which are accessible in medium-energy heavy-ion collisions \cite{HADES:2019auv}.)
Implications of the conjecture of a first-order phase transition emerging from a QCD critical endpoint
continuing to small temperatures and large baryon densities
\cite{Du:2024wjm}
can also be  studied by neutron--hybrid-quark stars.

In fact, a cold EoS 
encoding a strong local softening due to the first-order phase transition 
can give rise to the third island of compact stars. 
In special cases, these appear as twins of neutron stars 
\cite{Li:2024sft,Laskos-Patkos:2024fdp,Li:2022ivt}.

While the standard model  of particle physics (SM) seems to accommodate nearly all of the observed
phenomena of the micro-world, severe issues remain. Among them is the 
fundamental problem of the very nature of dark matter (DM): 
Astrophysical and cosmological observations seem to require inevitably its existence, 
but details remain elusive despite many concerted attempts.
Supposed DM behaves like massive particles,
these could be captured gravitationally in the centers of compact stars 
\cite{Karkevandi:2021ygv,Dengler:2021qcq,Hippert:2022snq},
thus providing a non-SM component there.
This would be an uncertainty on top of the less reliably known SM-matter state
at supra-nuclear densities.
Beyond the SM, also other feebly interacting particles could populate compact stars. A candidate scenario is
provided, for instance, by mirror world 
\cite{Alizzi:2021vyc,Goldman:2019dbq,Beradze:2019yyp,Berezhiani:2021src,Berezhiani:2020zck}.
There are many proposals of portals from our SM-world to such beyond-SM scenarios,
cf.\ \cite{Beacham:2019nyx,Pitz:2024xvh,Yang:2024ycl,Diedrichs:2023trk},
which can be tested by their consequences for compact stars.
New mass-radius data \cite{Kini:2024ggu} point to a puzzling compact object,  XTE J1814-338,
which stimulates various explanations \cite{Pitz:2024xvh,Yang:2024ycl,Laskos-Patkos:2024fdp}.
Also  HESS J1731-347 \cite{Doroshenko:2022nwp} seems to be a particularly interesting object 
\cite{Li:2024sft,Li:2022ivt}.
We confront these and other data with our core-corona decomposition (CCD).

Our note is organized as follows.
Section \ref{sect:II} is devoted to the CCD, where a specific EoS is deployed
for the explicit construction. 
The comparison of the CCD with data is presented in Section \ref{sect:III},
where we include XTE J1814-338 as compact object with a large massive core
(which may contain a DM component or/and a special SM material component)
and speculate whether HESS J1731-347 belongs to the sequence of SM matter neutron stars
with radii of about 12~km.
We summarize in Section \ref{sect:summary}. 
Appendix \ref{app:A} lists some entries to data,
and Appendix  \ref{app:B} is devoted to a brief comment on a particular core model
based on a statistically determined EoS from multimessenger data; a remark refers to the
usefulness of the CCD when dealing with a first-order phase transition. 

\section{Core-corona decomposition}\label{sect:II}

The standard modeling of compact star configurations is based on 
the Tolman-Oppenheimer-Volkoff (TOV) equations
\begin{align}
\frac{\mbox{d} p(r)}{\mbox{d} r } &= 
- G_N \frac{[e(r)+p(r)] [m(r)  + 4 \pi r^3 p(r)]}{r^2 [1 - 2 G_N \frac{m(r)}{r}]}, \label{eq:p_prime} \\
\frac{\mbox{d} m(r)}{\mbox{d} r} &= 4 \pi r^2 e(r), \label{eq:m_prime}
\end{align}  
resulting from the energy-momentum tensor of a static isotropic fluid 
(described locally by pressure $p$ and energy density $e$, solely relevant for the medium) 
and spherical symmetry of both space-time and matter, 
within the framework of Einstein gravity without cosmological term \cite{Schaffner-Bielich:2020psc}.
Newton's constant is denoted by $G_N$, and
natural units with $c = 1$ are used,
unless when relating mass, length, pressure and energy density, where $\hbar c$ is needed.

Given a unique relationship of pressure $p$ and energy density $e$ as EoS
$e(p)$, in particular at zero temperature, the TOV equations are integrated customarily with boundary conditions
$p(r) = p_c - {\cal O} (r^2)$ and $m(r) = 0 + {\cal O} (r^3)$ at small radii $r$,
and $p(R) = 0$ and $m(R) = M$ with $R$ as circumferential radius and $M$ as gravitational mass
(acting as parameter in the external (vacuum) Schwarzschild solution at $r > R$).
The quantity $p_c$ is the central pressure.
The solutions $R(p_c)$ and $M(p_c)$
provide the mass-radius relation in parametric form $M(R)$, being a curve.

Many concerted efforts, e.g.\ \cite{Rutherford:2024srk} and further citations therein,
are spent to pin down the EoS from observational data, including the
mass and radius values and dynamics of merging binary compact objects and their gravitational waves,
with significant input from nuclear physics and relativistic heavy-ion physics.
In fact, big deal of efforts is presently concerned about the EoS at supra-nuclear densities \cite{Reed:2021nqk}.
For instance,
Fig.~1 in \cite{Annala:2019puf} exhibits the recently admitted uncertainty: up to a factor of ten in pressure 
as a function of energy density. At asymptotically large energy density, perturbative QCD constraints
the EoS, though it is just the non-asymptotic supra-nuclear density region which determines crucially
the maximum mass and
whether twin stars may exist or quark-matter cores appear in neutron stars. 
Accordingly, one can fill this gap by a huge number of test EoSs 
to scan through the possibly resulting manifold of mass-radius curves, see 
\cite{Altiparmak:2022bke,Ayriyan:2021prr,Greif:2020pju}.
However, as stressed above,
the possibility that neutron stars may accommodate other components than
SM matter, e.g.\ exotic material as DM
\cite{Anzuini:2021lnv,Bell:2019pyc,Das:2021hnk,Das:2021yny},
can be an obstacle for the safe theoretical modeling of a concise mass-radius relation in such a manner. Of course, 
inverting the posed problem with sufficiently precise data of masses and radii as input offers 
a promising avenue towards determining the EoS 
\cite{Newton:2021yru,Huth:2021bsp,Ayriyan:2021prr,Blaschke:2020qqj,Raaijmakers:2021uju,Raaijmakers:2019dks,Raaijmakers:2019qny}.

In contrast, if the EoS $p(e, \mathfrak{x})$ depends on some additional parameter $\mathfrak{x}$ 
(e.g.\ for the composition of the matter, may it be
a strangeness fraction \cite{Yang:2024ycl}
or temperature \cite{Carlomagno:2024vvr}),
a bundle $M(R, \mathfrak{x}=const)$ curves 
is generated by solving the TOV equations for various fixed values of $\mathfrak{x}$.
Smooth variation of $\mathfrak{x}$ generates an area $M(R, \mathfrak{x})$.
Analogously, treating a DM component as a second fluid which only interacts with the SM component
via the common gravitational field, one needs a second EoS, $p^\mathrm{DM}(e^\mathrm{DM})$, and the
two-fluid TOV equations deliver $M(p_c, p_c^\mathrm{DM})$ and  
$R(p_c, p_c^\mathrm{DM})$, meaning again an area in the mass-radius plane.
Besides the still pertinent uncertainty in $p(e)$, it is the wide range of possible DM scenarios  
(fermionic or bosonic or mirror DM) which
make the modeled mass-radius relations rather indefinite. The combinations of multimessenger data
including the tidal deformation and specific gravitational wave forms in time have the potential to
constrain the scenarios. 

Here, we pursue another perspective \cite{Zollner:2022dst,Zollner:2023myk}.
We parameterize the supra-nuclear core by a radius~$r_x$ and the
included mass~$m_x$ and integrate the above TOV equations only within the corona,\footnote{Our notion 
``corona" is a synonym for  ``mantel" or ``crust" or ``envelope" or ``shell" or ``halo".
It refers to the complete part of the compact star outside the core, $r_x \le r \le R$.} 
i.e.\ from pressure $p_x$
to the surface, where $p = 0$. This yields the total mass $M(r_x, m_x; p_x)$ and the total radius $R(r_x, m_x; p_x)$
by assuming that the corona EoS $p(e)$ is reliably known at $p \le p_x$
and only SM matter occupies the region $r \ge r_x$. 
Clearly, without any knowledge
of the matter composition at $p > p_x$ (may it be SM matter with an uncertainly known EoS or
may it contain a DM admixture, for instance, or  monopoles or some other type of
``exotic" matter) one does not get a simple mass-radius relation by such a procedure, 
but admissible area(s) over the mass-radius plane, depending on the core
parameters $r_x$ and $m_x$ and the matching pressure $p_x$ and related energy density $e_x$.
This is the price of avoiding a specified model of the core matter composition.
However, the CCD is a simple and efficient approach to quantify 
the appearance of some ``exotics"
by a displacement from the mass-radius curve related to a SM matter EoS. 
For a SM matter-only core with EoS $p(e) \approx p^\mathrm{SM} (e)$, the core parameters are strongly correlated,
$m_x(p_c)$$, r_x(p_c)$ for $p_c \ge p_x$, thus yielding masses and radii near to or on the
$M(R)$ curve provided by  $p^\mathrm{SM} (e)$.
The limits $r_x \to 0$, $m_x \to 0$ and $p_x \to p_c$ turn the CCD
in the conventional one-fluid TOV equations. 

If the core is occupied by a one-component SM medium, the region 
$p > p_x$ and $e > e_x$ can be mapped out by many test EoSs 
which obey locally the causality constraint 
to obtain the corresponding region in the mass-radius plane, cf.\  Fig.~2 in \cite{Gorda:2022jvk} for an example
processed by Bayesian inference.
This is equivalent, to some extent, to our CCD for SM matter-only.

Note that our core-corona approach relies on the assumption that
the region $r \in [r_x, R]$ is occupied only by SM matter with a trustable EoS.
Thus, scenarios as in \cite{Pitz:2024xvh}, where bosonic DM forms a halo around
a core with SM + DM components, are not accessible by our CCD.

\section{Comparing CCD with NICER and other data}\label{sect:III}

\subsection{Data selection}\label{subsect:data}

To demonstrate CCD features,
we select four representative data sets out of the multitude of published values:
$(R_\mathrm{km}, M/M_\odot)$ =
($7.0^{+0.4}_{-0.4}$,  $1.21^{+0.05}_{-0.05}$)  for XTE J1814-338 \cite{Kini:2024ggu},
($10.4^{+0.86}_{-0.78}$,  $0.77^{+0.20}_{-0.17}$)\footnote{The very small mass value is scrutinized
in \cite{Salmi:2024bss}.
}  
for HESS J1731-347  \cite{Doroshenko:2022nwp},
($11.36^{+0.95}_{-0.63}$,   $1.42^{+0.04}_{-0.04}$)${}^\mathrm{NICER}$ for PSR J0437-4715 
\cite{Choudhury:2024xbk} 
and
($12.49^{+1.28}_{-0.88}$, $2.073^{+0.069}_{-0.069}$)${}^\mathrm{NICER}$   for PSR J0740+6620 
\cite{Salmi:2024aum}.
These data are exhibited in Fig.~\ref{fig:1}.
More (NICER) data are mentioned in Appendix \ref{app:A}.
For our goal, such complementing information is not needed.

\subsection{CCD compared to data}\label{subsect:CCD}

The solid curve in Fig.~1 exhibits an example of the SM matter EoS NY$\Delta$ \cite{Li:2019fqe}.
Basically, it needs to be supplemented by a proper crust EoS. For the sake of simplicity, however,
we extrapolate linearly the tabulated lowest pressure point at $(p_1, e_1) = (8.19, 216.92)~\mathrm{MeV/fm}^3$ 
to $p = 0$ for a chosen value of $e_0$.
The solid and faint dashed and dotted curves in Fig.~1 are based on $e_0 = 1, 100$, and $0.1~\mathrm{MeV/fm}^3$,
respectively,
thus displaying schematically the dependence of the $M(R)$ curve on the crust details.
The asterisks on the solid curve are for central pressures $p_c = 150, 100$ and $50~\mathrm{MeV/fm}^3$ 
(from top to bottom), respectively. 
Note that rotational effects are assumed to be subleading.

\begin{figure}[ht]
\includegraphics[width=0.66\columnwidth]{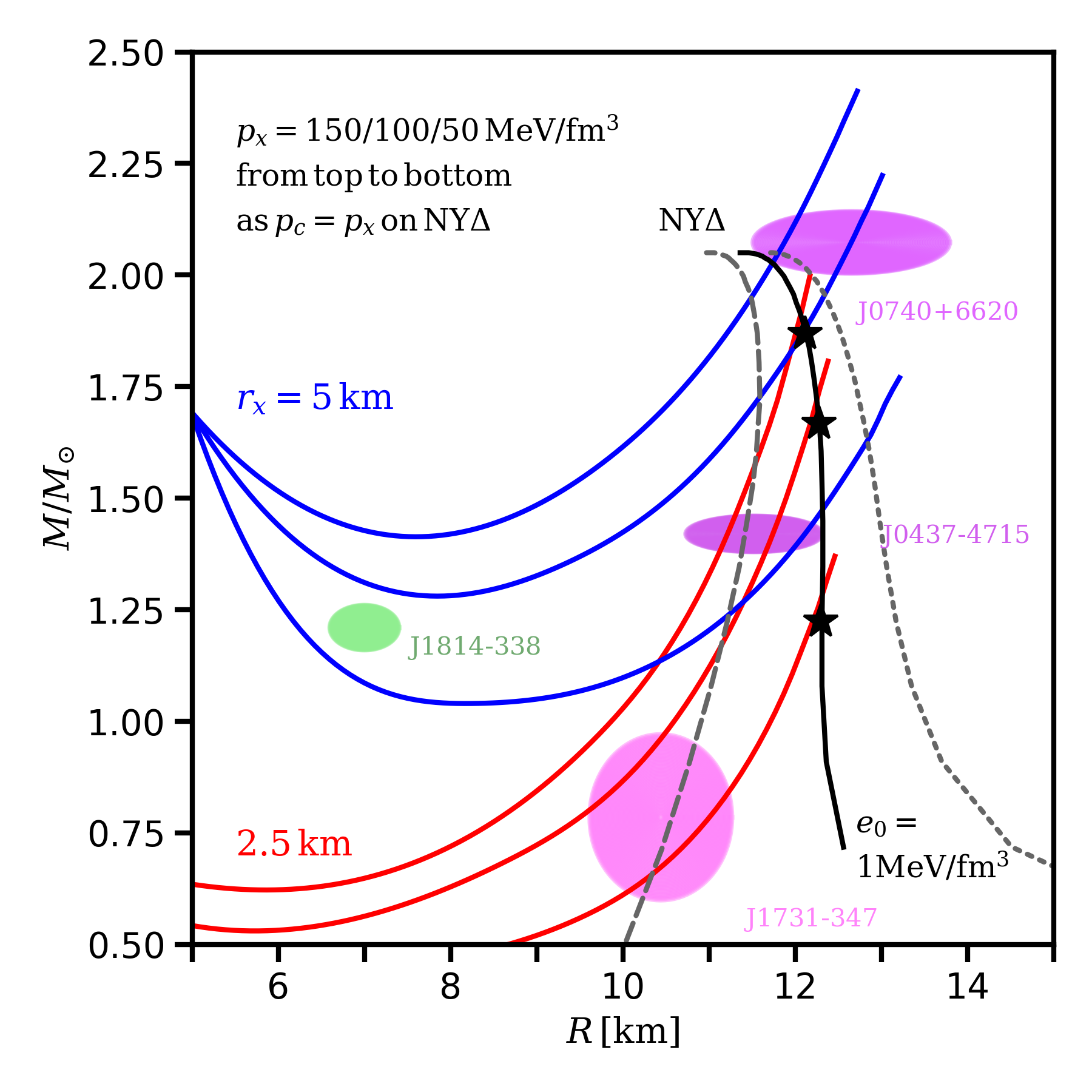}
\caption{Mass-radius relation of compact (neutron) stars in the CCD
with core radii $r_x = 2.5$~km (red curves) and 5~km (blue curves)
and varying core masses $m_x$ from small values (right side) to large values (left side, terminating
at $2 G_N m_x/r_x \to 1$).
The pressures at the core radii $r_x^+$ are 
$p_x = 150$ (top), 100 (middle) and 50~$\mathrm{MeV/fm}^3$ (bottom).
The solid black curve is for the EoS NY$\Delta$ \cite{Li:2019fqe} with linear interpolation
down to a minimum energy $e_0 = 1~\mathrm{MeV/fm}^3$ at $p = 0$.
The asterisks are for central pressures $p_c = 150, 100$ and  $50~\mathrm{MeV/fm}^3$ 
(from top to bottom). The faint dashed (dotted) curve employs
$e_0 = 100~\mathrm{MeV/fm}^3$ ($0.1~\mathrm{MeV/fm}^3$)
showing schematically the impact of the surface crust.
The displayed data (colored ellipses) are listed in Subsection \ref{subsect:data}.
\label{fig:1} 
}
\end{figure}

Choosing core radii of $r_x = 2.5$ and 5~km, one gets the red and blue curves
for running values of the core masses $m_x$. 
Each curve depends on $p_x$; our selected values are 150, 100 and 50~$\mathrm{MeV/fm}^3$
(from top to bottom). 
A core radius of about 5~km and core mass of about $1.07~M_\odot$ and
$p_x \approx 50~\mathrm{MeV/fm}^3$ match well the data of the outlier XTE J1814-338. 

While NICER data seem to line up at $R \approx 12~\mathrm{km}$,
the extraordinarily light object HESS J1731-347 points to a somewhat smaller radius,
although not incompatible with the mean value of $R \approx 12~\mathrm{km}$.
If really displaced from the mean radius, again the CCD would favor
a small core radius of about $2.5~\mathrm{km}$ and core mass of $0.1745~M_\odot$, with the
same value of $p_x \approx 50~\mathrm{MeV/fm}^3$,
which corresponds -- for NY$\Delta$ -- to a conformality measure $\Delta_x := \frac13 - \frac{p_x}{e_x} \approx 0.21$
as advocated in \cite{Marczenko:2023txe} at $e_x = 400~\mathrm{MeV/fm}^3$.
Different evolutionary stages w.r.t.\ accretion of the DM could be at the origin
of the difference in core radii and masses.

\subsection{Pressure and mass profiles in the corona}\label{subsect:profiles}

The scale settings 
$p = p_x \tilde p$,
$m = m_x \tilde m$,
$r = r_x \tilde r$ and
$e = e_x \tilde e$
turn the TOV equations into dimensionless form,
to be integrated from $\tilde p = \tilde m = \tilde r = \tilde e = 1$
to the surface, where $\tilde p = 0$.
The three parameters $(r_x, m_x; p_x)$ translate into 
$(F_1 := G_N m_x / r_x, \, F_2 := r_x^3 p_x  / m_x; \, e_x/p_x)$, thus not allowing
for a simple scaling or design of master curves $\tilde p (\tilde r)$ and $\tilde m (\tilde r)$ depending only
on one or on two parameters, e.g.\ $F_{1,2}$. Thus, we return to unscaled quantities.

\begin{figure}[t!]
\includegraphics[width=0.49\columnwidth]{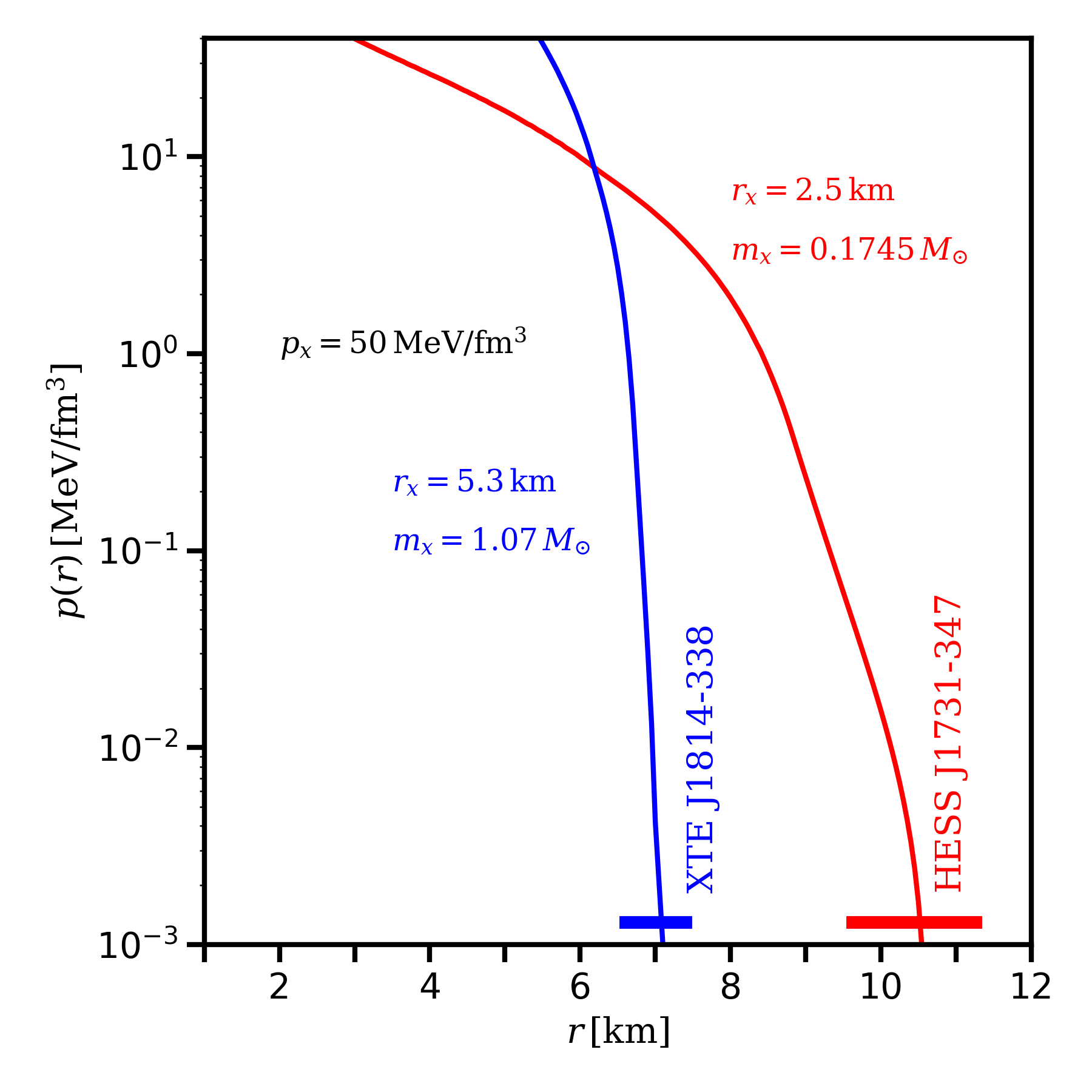} \hspace{-6mm}
\includegraphics[width=0.49\columnwidth]{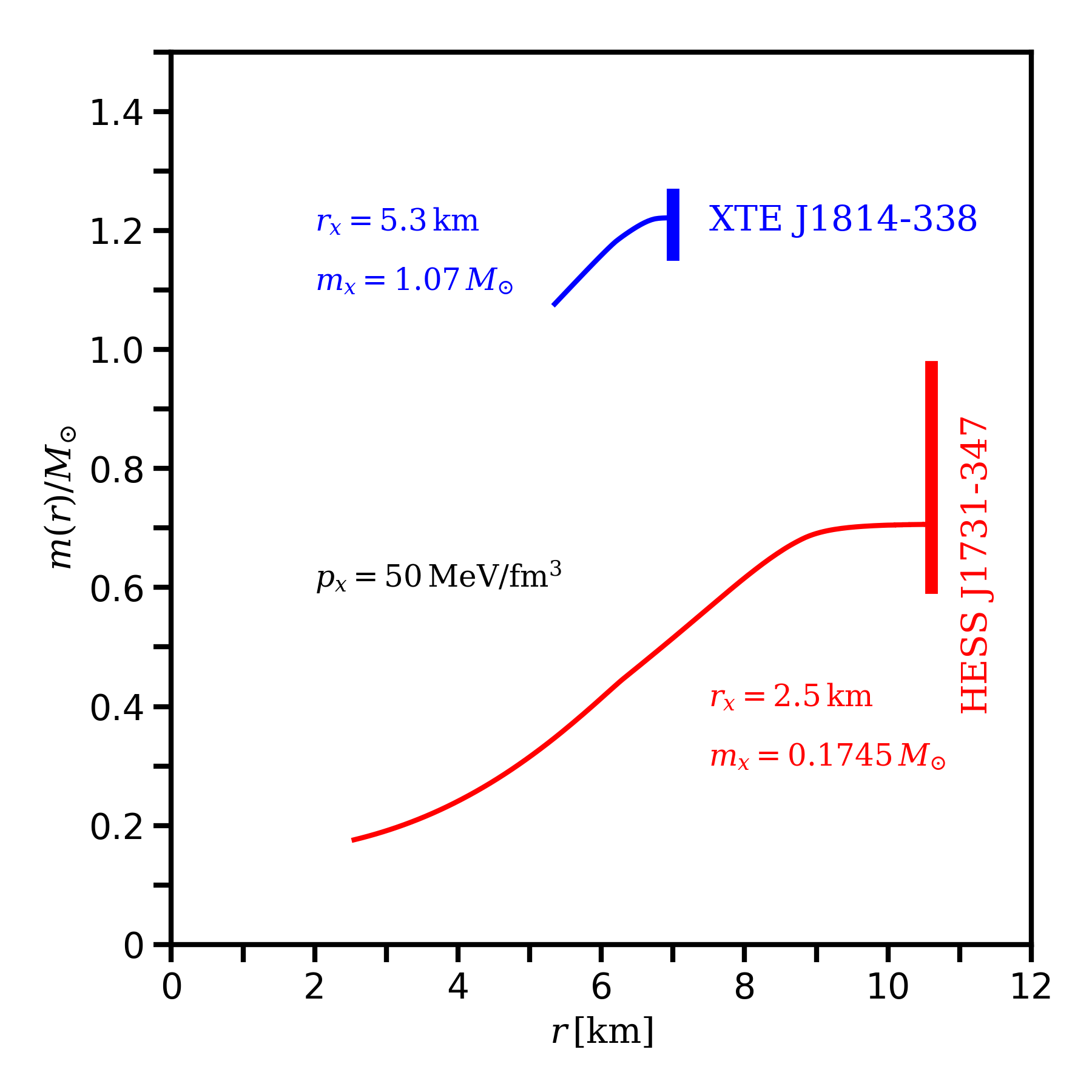}
\caption{Profiles of pressure $p (r)$ (left panel) and mass $m (r)$ (right panel) in the corona
with common value $p_x = 50~\mathrm{MeV/fm}^3$. 
The values
$(r_x, m_x) = (2.5~\mathrm{km},  0.1745~M_\odot)$ (in red)
and
$(5.3~\mathrm{km}, 1.07~M_\odot)$ (in blue)
match the current data of J1814-338 and J1731-347 (see Subsection \ref{subsect:data} and Appendix \ref{app:A})
which are depicted  by bars.
The EoS is NY$\Delta$ as in Fig.~\ref{fig:1}.
\label{fig:2} 
}
\end{figure}

The respective (unscaled)  profiles $p(r)$ und $m (r)$ in the corona 
with EoS NY$\Delta$ are exhibited in Fig.~\ref{fig:2}.
The CCD requires a fairly large and massive core to match the XTE J1814-338 data
leaving less space for the corona. In contrast, the HESS J1731-347 data are nicely reproduced by
a small  low-mass core. 

While the accomplished decomposition seems to leave the determination of the advocated core by explicitly
accommodating either a SM matter EoS or a SM + DM mixture with separate EoSs such to reproduce the
triple $(r_x, m_x; p_x)$, our construction is not yet universal since it depends on the actually employed
corona EoS. Admittedly, one has to test the robustness of the deduced values $(r_x, m_x; p_x)$
by using others than the NY$\Delta$ EoS, in particular by updated outer crust models, cf.\ \cite{Suleiman:2021hre}.

In some sense, our results support the model of a two-family approach
based on two distinct classes of EoSs \cite{DiClemente:2021dmz}. 
 
\section{Summary} \label{sect:summary}

The core-corona decomposition (CCD) relies on the agnostic assumption that the EoS of compact (neutron) star matter 
(i) is reliably known up to energy density $e_x$ and pressure $p_x$ and 
(ii) standard model (SM) matter occupies the star as the only component at radii $r > r_x$. 
The base line for static, spherically symmetric configurations is then provided by the TOV equations,
which are integrated, for $r \in [r_x, R]$, to find the circumferential radius $R$ (where $p(R) = 0$) and the
gravitational mass $M = m(R)$. We call the region $r \in [r_x, R]$ ``corona", 
but ``crust" or "mantle" or ``envelope" or ``shell" or ``halo"
are also suitable synonyms. The region $r \in [0, r_x]$ is the ``core", parameterized by 
the included mass $m_x$. The core must support the corona pressure at the interface, i.e.\ $p(r_x^-) = p(r_x^+)$,
assuming either pure SM matter or a dark matter (DM) component  feebly interacting with the SM matter component. 
The core can contain any material compatible with the symmetry requirements.
In particular, it could be modeled by multi-component fluids with SM matter plus DM
or anything else beyond the SM. 
Alternatively, a first-order phase transition (FOPT) could be accommodated in the SM matter-only one-fluid core.
Then, $p_x = p_\mathrm{FOPT}$ would be appropriate, see Appendix \ref{app:B}.

The currently available mass-radius data of XTE J1814-338 and HESS J1731-347
point to an averaged core mass density
$\langle \rho \rangle_x := 3 m_x /(4 \pi r_x^3) \approx 3 \times 10^{15}~\mathrm{g/cm}^3$,
with HESS J1731-347 ($r_x \approx 2.5~\mathrm{km}$, $m_x \approx 0.17~M_\odot$)
marginally displaced from the mean radius value of about 12~km,
while XTE J1814-338 ($r_x \approx 5.3~\mathrm{km}$, $m_x \approx 1.07~M_\odot$)
seems to belong to another class of very compact (neutron) stars. 

The tidal deformability and stability properties remain as challenging issues. 
Improved data will provide further constraints and pave the way of explicating the core properties
w.r.t.\ the options of a FOPT and/or DM admixtures. 

\begin{appendix}

\section{Data of radii and masses}\label{app:A}

Ordered tentatively with increasing radii, 
one can group the data of 
XTE J1914-338 \cite{Kini:2024ggu},
HESS J1731-347  \cite{Doroshenko:2022nwp},
PSR J0437-4715  \cite{Choudhury:2024xbk,Pang:2021jta,Miller:2021qha},
PSR J1231-1411 \cite{Salmi:2024bss},
PSR J0030+0451 \cite{Riley:2019yda,Miller:2019cac,Vinciguerra:2023qxq},
PSR J0740+6620 \cite{Riley:2021pdl,Fonseca:2021wxt,Salmi:2024aum,Miller:2021qha,Pang:2021jta}, 
in two categories: (i) such ones with radii $R \approx (12 \pm 1)~\mathrm{km}$
and (ii) $R < 11~\mathrm{km}$.    
Further information stems from GW190814: gravitational waves from the coalescence of a 23 solar mass black hole 
with a $2.6 M_\odot$  compact object \cite{LIGOScientific:2020zkf}.
The black widow pulsar PSR J0952-0607 also points to a large mass of $2.35^{+0.17}_{-0.17}~M_\odot$ 
\cite{Romani:2022jhd} with implications studies in \cite{Ecker:2022dlg}.
The kilonova GW170817 has been interpreted \cite{LIGOScientific:2018cki}
as merger event with both components having radii of
$R = 11.9^{+1.4}_{-1.4}~\mathrm{km}$ 
under the requirement that the EoS allows for neutron stars with masses larger than $1.97~M_\odot$, 
thus supporting the mean radius of about 12~km.
Indeed, \cite{Miller:2021qha}  quotes $R =12.45^{+0.65}_{-0.65}~\mathrm{km}$ for a $1.4~M_\odot$ neutron star and 
$R = 12.35^{+0.75}_{-0.75}~\mathrm{km}$ for a $2.08~M_\odot$ neutron star.

Despite some variation of the mass for objects in category (i)
(mostly $M \ge 1.4 M_\odot$, including in particular PSR J0030+0451),
a mean radius of about  12~km can be inferred. 
(The radius difference of J0437-4715 and J0030+0451 has been discussed w.r.t.\ a hint of twin stars \cite{Li:2024sft}.)
The small mass of HESS J1731-347 and the 
smaller radius of XTE J1814-338 
in category (ii) stimulates currently investigations 
\cite{Pitz:2024xvh,Yang:2024ycl,Laskos-Patkos:2024fdp} on their nature, e.g.\ whether DM admixtures 
could be at the origin of these abnormalities.
However, one should be aware of the model dependence 
(e.g.\ w.r.t.\ EoS inference techniques, background constraints, joint analyses)
of some of the fits yielding a notable spread of 
the quoted radii and masses.

\section{Considering a one-fluid core}\label{app:B}

To present an explicit example of an one-fluid core we adapt the EoS related to the QCD trace anomaly $\Delta$, i.e.\
$p(e) = e [\frac13 - \Delta(e)]$ and squared sound velocity
$v_s^2 = \frac13 - \Delta - e \frac{\partial \Delta}{\partial e}$.
In \cite{Marczenko:2023txe}, $\Delta (e)$ has been statistically determined
with constraints from multimessenger neutron star data w.r.t.\ hints of approaching conformality 
in the cores.\footnote{$\Delta(\eta)$ with $\eta := \ln (e/ 150~\mathrm{MeV/fm}^3)$,
originally proposed in \cite{Fujimoto:2022ohj}, agrees -- with a tiny correction of one parameter --
with $\Delta^{\mathrm{NY}\Delta} (\eta > 0.9)$ \cite{Li:2019fqe}, see Fig.~3-left in \cite{Zollner:2023myk}.
The plots of $\Delta (e)$ and $v_s^2 (e)$ in \cite{Marczenko:2023txe}, see Fig.~4 there, can be parameterized by
$\Delta = 0.33 \left[1-\frac{A}{1+\exp\{-\kappa (\eta - \eta_c) \} } \right]
+ G_G \exp\{- \frac{(\eta - \eta_G)^2}{2 \sigma_G} \}$        
with $\eta := \ln (e / \hat e_0)$,
$\hat e_0 = 0.12~\mathrm{GeV/fm}^3$,
$A = 1.52$,
$\kappa = 3.582856$,
$\eta_c = 1.357143$,
$G_G = 0.2926$,
$\eta_G = 4.045714$,
$\sigma_G = 2$.
\label{footnote.Delta}
}
The one-fluid TOV equations 
are integrated from $r = 0$,
where $p = p_c$ and $m = 0$,
to $r_x$, where $p (r_x) = p_x$ and $m_x = m (r_x)$.

\begin{figure}[t!]
\includegraphics[width=0.49\columnwidth]{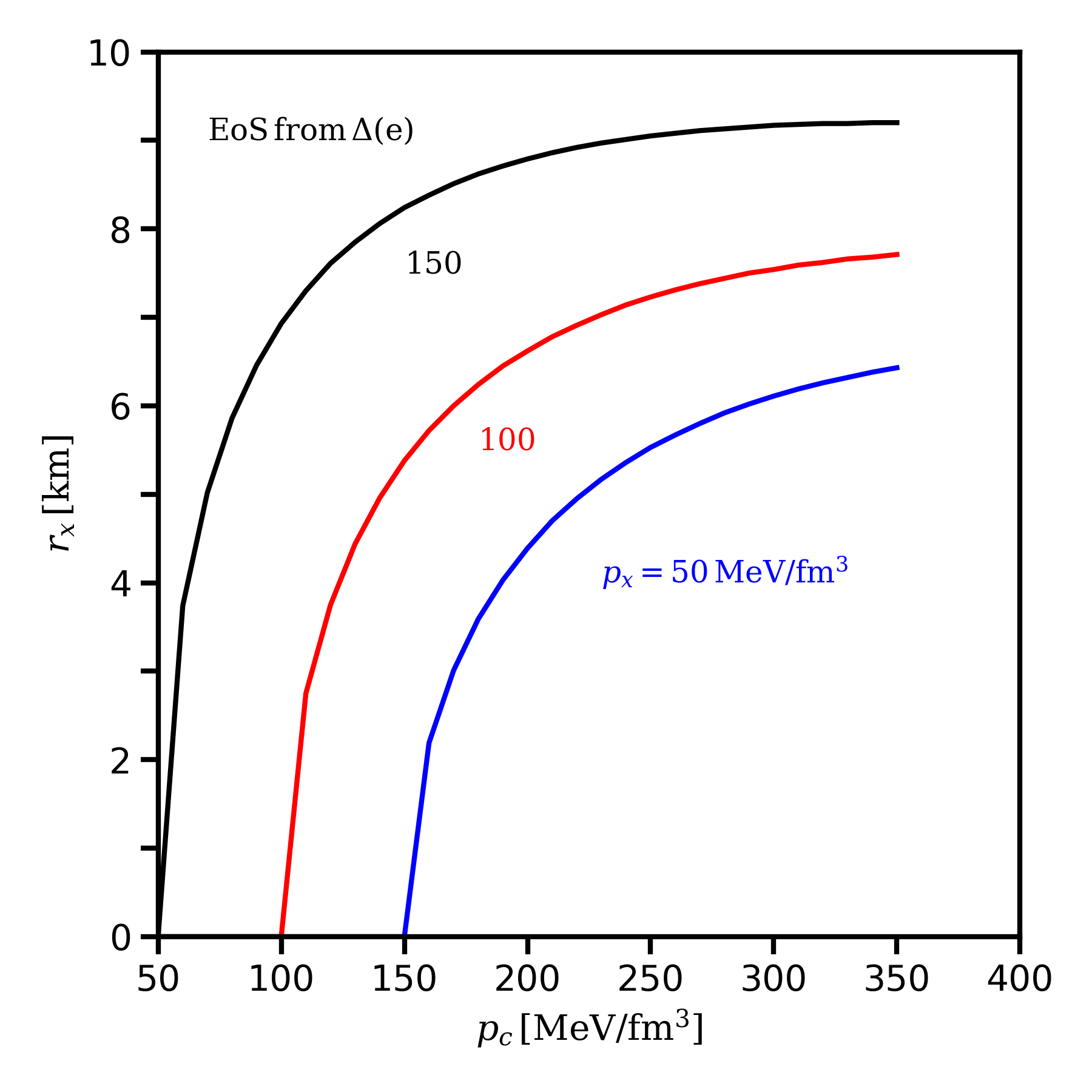}  \hspace{-6mm}
\includegraphics[width=0.49\columnwidth]{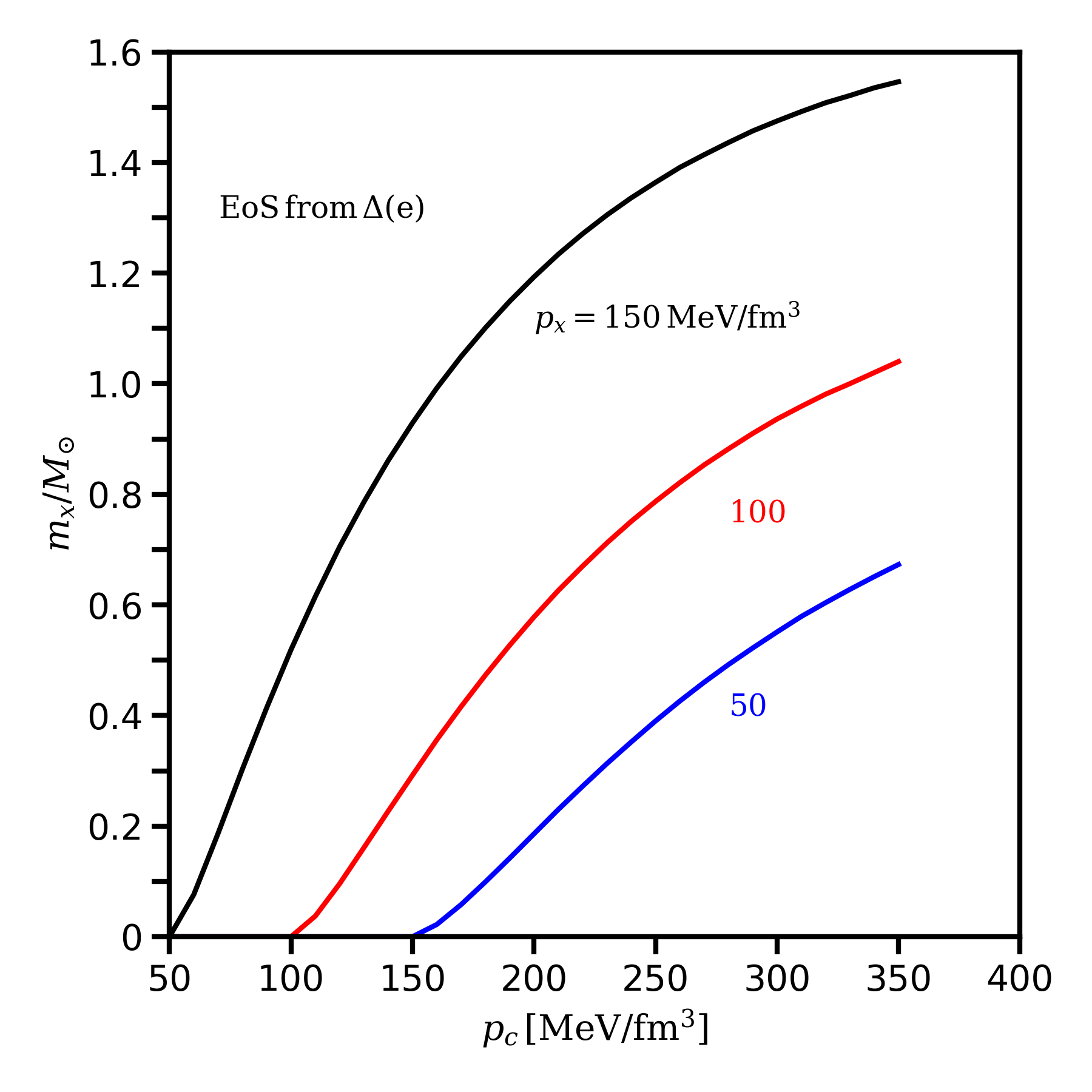}
\caption{Core radii $r_x$ (left panel) and masses $m_x$ (right panel) as a function of the central pressure $p_c$ for various 
values of the pressure $p_x = 50, \, 100$ and $150~\mathrm{MeV/fm}^3$ (from top to bottom)
at the core surface, where $p(r_x) = p_x$.
The EoS reads $p = e [\frac13 - \Delta(e)]$, where $\Delta(e)$ is a fit of the results in \cite{Marczenko:2023txe},
see footnote \ref{footnote.Delta}.
This example of a one-fluid core model does not deliver consistent values 
of $r_x(p_c; p_x)$ and $m_x(p_c; p_x)$ at $p_x = 50~\mathrm{MeV/fm}^3$
which are needed -- in combination with the corona model -- to match the XTE J1814-338 data \cite{Kini:2024ggu}.
\label{fig.Delta} 
}
\end{figure}

The emphasis is here on the non-trivial dependence of both
$\Delta (e)$ and  $v_s^2 (e)$ which encode the EoS.  
The corresponding core radii and masses 
as functions of the central energy density $e_c$ are exhibited in Fig.~\ref{fig.Delta}.
Both, core radii and masses increase with central pressure. 
For comparison, somewhat smaller values of $r_x(p_c)$ and $m_x(p_c)$
would be obtained when using NY$\Delta$ as core EoS.  
Note that, for a one-fluid core with given EoS and $p_x$, the values of $r_x$ and $m_x$ become
correlated due to the $p_c$ dependence, and the usual stability criteria apply.

The CCD is useful for splitting off details of the corona.
The particular case of an one-fluid SM matter core with conjectured EoS is handled  analogously 
as in Subsection \ref{subsect:profiles}, especially when dealing with a first-order phase transition (FOPT).\footnote{
The underlying EoS with FOPT is $p(e) = p_\mathrm{FOPT}$ for $e \in [e_x, e_2]$,
where -- at $e < e_x$ and $p <  p_\mathrm{FOPT}$ -- the corona EoS (e.g.\ NY$\Delta$ in our examples above) applies,
while  -- at $e > e_2$ and  $p >  p_\mathrm{FOPT}$ -- the core EoS (see footnote \ref{footnote.Delta} for an example) applies. 
Combining $p^\mathrm{core} (e)\vert_{p > p_x}$ and $p^\mathrm{corona} (e)\vert_{p< p_x}$ 
trivially provides a FOPT model at $p_x \equiv p_\mathrm{FOPT}$
if  $e^\mathrm{core} (p_x) > e^\mathrm{corona} (p_x)$. 
}

\end{appendix}

\begin{acknowledgements}

One of the authors (BK) acknowledges continuous discussions with
J.~Schaffner-Bielich and K.~Redlich for the encouragement to deal with the current topic.
Interesting discussions with A.~Junghans initiated the present note.
The work is supported in part by the European Union’s Horizon 2020 research
and innovation program STRONG-2020 under grant agreement No 824093. 
We apologize for not quoting all of the overwhelmingly many entries which contribute to the current topic.

\end{acknowledgements}

\end{document}